\documentclass[twocolumn,showpacs,amssymb,amsmath,aps,pra]{revtex4}
\newcommand{\half}{\mbox{$\frac{1}{2}$}}
\usepackage{graphics}
\usepackage{bm}
\begin{document}
\title{Entanglement between distant qubits in cyclic XX chains}
\author{N. Canosa, R. Rossignoli}
\affiliation{Departamento de F\'{\i}sica-IFLP,
Universidad Nacional de La Plata, C.C.67, La Plata (1900), Argentina}
\begin{abstract}
We evaluate the exact concurrence between any two spins in a cyclic $XX$ chain
of $n$ spins placed in a uniform transverse magnetic field, both at zero and
finite temperature, by means of the Jordan-Wigner transformation plus a number
parity projected statistics. It is shown that while at $T=0$ there is always
entanglement between any two spins in a narrow field interval before the
transition to the aligned state, at low but non-zero temperatures the
entanglement remains non-zero for arbitrarily high fields, for any pair
separation $L$, although its magnitude decreases exponentially with the field.
It is also demonstrated that all associated limit temperatures approach a
constant non-zero value in this limit, which decreases as $L^{-2}$ for $L\ll n$
but exhibit special finite size effects for distant qubits ($L\approx n/2$).
Related aspects such as the different behavior of even and odd
antiferromagnetic chains, the existence of $n$ ground state transitions and the
thermodynamic limit $n\rightarrow\infty$ are also discussed.
\end{abstract}
\maketitle

\section{Introduction}
Quantum entanglement denotes those correlations with no classical analogue that
can be exhibited by composite quantum systems and that constitute one of the
most fundamental features of quantum mechanics. It is considered an essential
resource in the field of quantum information \cite{NC.00}, where it plays a key
role in various quantum information processing tasks such as quantum
teleportation \cite{Be.93} and quantum cryptography \cite{Ek.91}. It is also
playing an increasingly important role in condensed matter physics, providing a
new perspective for understanding quantum phase transitions and collective
phenomena in strongly correlated systems \cite{ON.02,OS.02,V.03,T.04}.

In particular, there has been considerable interest in investigating
entanglement in quantum spin chains with Heisenberg interactions
\cite{LSM.61,S.99}, since they provide a scalable qubit representation apt for
quantum processing tasks \cite{DV.99,BB.04} which can be realized in diverse
physical systems. Studies of the pairwise entanglement in the Ising and $XY$
models \cite{ON.02,OS.02,GKBV.01} and in the isotropic Heisenberg model
\cite{ABV.01,Wa.02,W.04,AK.04} at zero and finite temperature and in a
transverse uniform field, as well as in diverse $XX$, $XY$ and $XYZ$ models for
two or a small number of qubits \cite{Wa.01,KS.02,CR.04,AK.05,CZ.05}, have been
made. An important result is that the entanglement range may remain finite at a
quantum phase transition, limited for instance to first and second neighbors in
the Ising model \cite{ON.02,OS.02}, in contrast with the behavior of the
correlation length, which diverges at these points. Global thermal entanglement
has also been studied \cite{RC.05}, showing that limit temperatures for
pairwise entanglement are lower bounds to those limiting entanglement between
global partitions. A fundamental result for finite systems is that there is
always a {\it finite} limit temperature for entanglement, since any mixed state
becomes completely separable if it is sufficiently close to the full random
state \cite{Z.98,G.02}.

In this work we analyze the entanglement between any two spins in a cyclic
chain with nearest neighbor $XX$ coupling in a transverse magnetic field
(control parameter) by means of an exact analytic treatment valid for any spin
number $n$ and pair separation $L$, based on the Jordan-Wigner mapping and the
use of number parity projected statistics for $T>0$. Recent studies in $XX$
chains have focused either chains with a small number of spins
\cite{Wa.01,CZ.05,Y.02}, where results were obtained through direct
diagonalization, or open chains at zero temperature and field \cite{W.04}. We
will show that the $XX$ model offers very interesting properties such as
entanglement between {\it any} pair (full range) in a finite field interval
just before the critical point at $T=0$, which subsists for {\it large fields}
at low but non-zero temperatures $T<T_L$. Moreover, limit temperatures $T_L$
approach a {\it non-zero limit} for large fields, for {\it all} separations
$L$. It also displays $n$ ground state transitions at analytic field values,
entailing a stepwise variation of the entanglement range suitable for its use
as an entanglement switch. Let us mention that $XX$ chains have also been
employed for entanglement teleportation \cite{Y.02}.

Section II describes the formalism for evaluating the exact concurrence between
arbitrary sites both at zero and finite temperature. Section III describes the
main physical results, including the ground state transitions and concurrence
both in ferro- and antiferromagnetic systems, and a detail study of the limit
temperatures for entanglement. Conclusions are drawn in IV.

\section{Formalism}
We consider a cyclic chain of $n$ spins with nearest neighbor $XX$ coupling.
The Hamiltonian reads
\begin{subequations}\label{H}
\begin{eqnarray}
H&=&bS^z-v\sum_{j=1}^n(s^x_j s^x_{j+1}+s^y_{j}s^y_{j+1})\label{Ha}\\
&=&bS^z-\half v\sum_{j=1}^n(s^+_j s^-_{j+1}+s^-_{j+1}s^+_{j})\,,
\label{Hb}\end{eqnarray}
\end{subequations}
where $s^{x,y,z}_j$ are the spin components (in
units of $\hbar$) at site $j$, $s^{\pm}_j=s^x_j\pm is^y_j$, $S^z=\sum_{j=1}^n
s^z_j$ is the total spin along the direction of the transverse magnetic field
$b$ and $n+1\equiv n$. Our aim is to examine the entanglement between qubits at
arbitrary sites $i,j$ ($i\neq j$) in the thermal state
\begin{equation}
\rho(T)=Z^{-1}\exp[-\beta H]\,,\,\,\;\;\beta=1/T \,,
 \label{rho}\end{equation}
where $Z={\rm Tr}\exp[-\beta H]$ and $T$ is the temperature (we set Boltzmann
constant $k=1$). This entanglement is determined by the reduced pair density
$\rho_{ij}={\rm Tr}_{n-\{ij\}}\rho(T)$ and can be measured through the {\it
concurrence} \cite{W.98}
\begin{equation}
C_{ij}=[2\lambda_{M}-{\rm tr}\,R]_+\,,
\;\;\;\;R=\sqrt{\rho_{ij}^{1/2}\tilde{\rho}_{ij}\rho_{ij}^{1/2}}\,,\label{Cij}
\end{equation}
where $[u]_+\equiv (u+|u|)/2$, $\lambda_{M}$ denotes the largest eigenvalue of
the hermitian matrix $R$ and
$\tilde{\rho}_{ij}=4s^y_is^y_{j}\rho_{ij}s^y_is^y_{j}$ is the spin flipped
density (${\rm tr}R$ is the fidelity \cite{NC.00} between $\tilde{\rho}_{ij}$
and $\rho_{ij}$). The entanglement of formation \cite{Be.96} of the pair is
$E_{ij}=-\sum_{\nu=\pm}q_\nu\log_2 q_\nu$, where $q_\pm=(1\pm
\sqrt{1-C_{ij}^2})/2$ and is just an increasing function of $C_{ij}$,with
$E_{ij}=C_{ij}=1$ ($0$) for a maximally entangled (separable) pair state.

Since $H$ commutes with $S^z$ and is invariant under translation and inversion,
$\rho_{ij}$ will commute with the pair spin component $S^z_{ij}=s^z_i+s^z_{j}$
and its elements will depend just on the separation $|i-j|$. Hence, in the
standard basis of $S^z_{ij}$ eigenstates, it must be of the form
\begin{equation}
\rho_{ij}=\left(\begin{array}{cccc}p^{+}_L&0&0&0\\0&p_L&\alpha_L&0\\0&\alpha_L
&p_L&0\\0&0&0&p^-_{L}\end{array}\right)\,,\;\;\;L=|i-j|\,,
 \label{rij}\end{equation}
where $p^+_L+2p_L+p^-_L=1$, $p_L^+-p_L^-=2\langle s^z_i\rangle$ and
\begin{eqnarray}
p^{+}_L&=&\langle(s^z_i+\half)(s^z_{j}+\half)\rangle \,,\;\; \alpha_L=\langle
s^+_is^-_{j}\rangle\,. \label{pml}
 \end{eqnarray}
Here $\langle O\rangle\equiv {\rm Tr}\,\rho(T) O$ denotes the thermal average
of $O$ and $\langle s^z_i\rangle=\langle S_z\rangle/n$ is the intensive
magnetization. $\rho_{ij}$ commutes as well with the total spin of the pair
$(S^{ij})^2={\bf S}^{ij}\cdot{\bf S}^{ij}$, its eigenstates being the standard
triplet states and singlet $|\!\!\uparrow\uparrow\rangle$,
$|\!\!\downarrow\downarrow\rangle$ and
($|\!\!\uparrow\downarrow\rangle\pm|\!\!\downarrow\uparrow\rangle)/\sqrt{2}$,
with eigenvalues $p^{\pm}_L$, $p_L\pm \alpha_L$. The pair entanglement is
obviously driven by the mixing coefficient $\alpha_L$. The concurrence
(\ref{Cij}) becomes
\begin{equation}
C_{L}=2\left[\,|\alpha_L|-\sqrt{p^+_Lp^-_L}\,\right]_+\,,
 \label{CL}\end{equation}
so that $\rho_{ij}$ is entangled if and only if $|\alpha_L|>\sqrt{p^+_Lp^-_L}$.
This condition also follows from the PPT criterion \cite{P.96}.

\subsection{Exact energy levels}
By means of the Jordan-Wigner transformation to fermion operators
$c^\dagger_{j}=s^+_j\exp[-i\pi\sum_{k=1}^{j-1}s^+_{k}s^{-}_{k}]$ \cite{LSM.61},
we may rewrite $H$ exactly as a bilinear form in $c^\dagger_j$, $c_j$ for {\it
each} value of the spin or fermion number parity
\[P\equiv \exp[i\pi N]\,,\;\;N=\sum_{j=1}^Nc^\dagger_jc_j=S^z+n/2\,.\]
The result for $P=\sigma=\pm 1$ is \cite{LSM.61}
\begin{eqnarray}
H_{\sigma}&=&\sum_{j=1}^n b(c^\dagger_jc_j-\half)-\half
v(1-\delta_{jn}\delta_{\sigma 1})
(c^\dagger_jc_{j+1}+c^\dagger_{j+1}c_j)\nonumber\\
&=&\sum_{k\in K_\sigma}\lambda_k (c'^\dagger_{k}c'_{k}-\half),
\;\;\;\;\;\lambda_k=b-v\cos\omega_k\,,\label{lk}
 \end{eqnarray}
where the fermion operators $c'^\dagger_{k}$ are related to $c^\dagger_j$ by a
parity dependent discrete Fourier transform
\begin{eqnarray} c^\dagger_j&=&{\frac{1}{\sqrt{n}}}
\sum_{k\in K_\sigma}e^{i\omega_k j}c'^\dagger_{k},\;\;\omega_{k}=2\pi k/n\,,\\
K_\sigma&=&\{-[\half n]+\half\delta_{\sigma 1},\ldots,[\half(n-1)]+
\half\delta_{\sigma 1}\}
 \end{eqnarray}
with $[\ldots]$ denoting integer part. The index $k$ is then {\it half-integer
(integer)} for $\sigma=1$ ($-1$).

The $2^n$ energies are then $\sum_{k\in K_\sigma}(N_k-1/2)\lambda_k$, where
$N_k=0,1$ and $\sigma=(-1)^{\sum_k N_k}$. Note that the single fermion energies
$\lambda_k$ depend on the global parity $\sigma$ and are degenerate
($\lambda_k=\lambda_{-k}$) for $|k|\neq 0,n/2$. It is also apparent from
(\ref{lk}) that the spectrum of $H$ is independent of the sign of $b$, and for
{\it even} $n$ {\it also of the sign of $v$}, as
$\cos\omega_{k'}=-\cos\omega_k$ for $k'=n/2-k$ and $k'$ belongs to the same
parity as $k$ if $n$ is even. This is also evident from (\ref{H}), since for
even $n$ the sign of $v$ can be inverted by a local transformation $s^{x,y}_j
\rightarrow (-1)^j s^{x,y}_j$ (and that of $b$ by $s^{y,z}_j\rightarrow
-s^{y,z}_j$). The concurrence (\ref{CL}) will then exhibit the same properties,
depending just on $|b|$ and for even $n$ just on $|v|$ \cite{Wa.01}.

\subsection{Exact partition function and concurrence}
The partition function $Z$ of the system is to be evaluated in the full
grand-canonical (GC) ensemble of the fermionic representation. However, due to
the parity dependence of the latter, this requires a {\it number parity
projected statistics} \cite{RCR.98}.  $Z$ can then be written as a sum of
partition functions for each parity,
\begin{eqnarray} \!\!\!\!\!\!Z&=&
{\rm Tr}\!\sum_{\sigma=\pm 1}\half(1+\sigma P)e^{-\beta H_{\sigma}}=
\half\! \sum_{\sigma=\pm 1} (Z^\sigma_0+\sigma Z^\sigma_1)\,,
 \label{Zp}\end{eqnarray}
where $\half(1+\sigma P)$ is the projector onto parity $\sigma$ and
\begin{eqnarray}
\!\!\!\!\!\!\!\!\!\!\!\!Z_\nu^\sigma&=&{\rm Tr}\,P^\nu e^{-\beta
H_{\sigma}}=e^{\beta b n/2}\!\!\! \prod_{k\in K_\sigma}(1+(-1)^\nu
e^{-\beta\lambda_k})
 \label{Znu}\,,\end{eqnarray}
for $\nu=0,1$. The expectation value of an operator $O$ can then be similarly
expressed as
\begin{eqnarray}
\langle O\rangle&=&\half Z^{-1}\sum_{\sigma=\pm 1}
(Z^\sigma_0\langle O\rangle_0^\sigma+\sigma Z^\sigma_1\langle
O\rangle_1^\sigma)\,,\label{Om}\\
\langle O\rangle_{\nu}^{\sigma}&=&(Z_\nu^\sigma)^{-1}
 {\rm Tr}\,[P^{\nu}e^{-\beta H_{\sigma}}O]\,,\;\;\; \nu=0,1\,.\label{Omnu}
\end{eqnarray}
In the case of many-body fermion operators, the thermal version of Wick's
theorem \cite{Ga.60} {\it cannot} be applied in the final average (\ref{Om}),
but it {\it can} be used for evaluating the partial averages (\ref{Omnu}) (as
$P^\nu e^{-\beta H_\sigma}=e^{-\beta H_\sigma+i\nu\pi N}$ is still the
exponential of a one-body operator), in terms of the contractions
\begin{eqnarray}
g_L\equiv\langle {c}^\dagger_i c_{j}\rangle_\nu^{\sigma}= \frac{1}{n}\sum_{k\in
K_{\sigma}} \langle c'^\dagger_kc'_k\rangle^{\sigma}_{\nu}\cos(L\omega_k)\,,
\end{eqnarray}
where $\langle {c'}^\dagger_k c'_{k}\rangle_\nu^{\sigma}=[1+(-1)^\nu
e^{\beta\lambda_k}]^{-1}$ (Eq.\ \ref{Omnu}). As $s^z_i=c^\dagger_ic_i-\half$,
this leads to
\begin{equation}
\langle s^z_i\rangle^\nu_\sigma=g_0-\half,\;\;
\langle(s^z_i+\half)(s^z_{j}+\half)\rangle_\nu^\sigma=g_0^2-g_L^2\,.
\label{wck2}
\end{equation}
Using the identity
$s_i^+s_{j}^-=s_i^+[\prod_{k=i+1}^{j-1}(s_k^+s_k^-+s_k^-s_k^+)]s_{j}^-$ for
$i<j$,  with $s_j^+s_{j+1}^+= c^\dagger_j c^\dagger_{j+1}$, $s_j^+s_{j+1}^-=
c^\dagger_jc_{j+1}$ \cite{LSM.61}, one also obtains
\begin{equation}\langle s^+_is^-_{j}\rangle_\nu^\sigma=\half{\rm Det}(A_L)\,,
 \label{wck3}\end{equation}
where $A_L$ is the $L\times L$ matrix of elements
\begin{equation}(A_L)_{ij}=2g_{i-j+1}-\delta_{i,j-1}\,,\label{Al}\end{equation}
i.e., ${\rm Det}(A_1)=2g_1$, ${\rm Det}(A_2)=4[g_1^2-g_2(g_0-\half)]$. All
terms in (\ref{rij}) and (\ref{CL}) can then be exactly evaluated.

In the thermodynamic limit $n\rightarrow\infty$, and for {\it finite} $L\ll n$,
we can ignore parity effects and replace sums over $k$ by integrals over
$\omega\equiv \omega_k$. We can then directly employ Wick's theorem in terms of
the elements
\begin{equation}
g_L=\langle c^\dagger_ic_j\rangle=\frac{1}{\pi}\int_{0}^\pi
\frac{\cos(L\omega)}{1+e^{\beta(b-v\cos \omega)}}d\omega\,.
 \label{cijinf}\end{equation}
This leads to (Eq.\ \ref{pml})
\begin{equation}p_L^+=g_0^2-g_L^2,\;\;\;\;\alpha_L=\half {\rm Det}(A_L)\,,
\label{wck}
\end{equation}
and $p_L^-=p_L^++1-2g_0$, where $A_L$ is constructed with the elements
(\ref{cijinf}). We then obtain the final expression
\begin{equation}
C_{L}=\left[|{\rm
Det}(A_L)|-2\sqrt{(g_0^2-g_L^2)((1\!-\!g_0)^2-g_L^2)}\right]_+\,.
 \label{Cex}\end{equation}
Note that for $T\rightarrow 0$,  Eq.\ (\ref{cijinf}) yields $g_L=0$ for $b>|v|$
and $g_L=\sin(L\omega)/(L\pi)$ (with $g_0=\omega/\pi$) for $|b|<|v|$, where
$\cos\omega=b/|v|$.

When the ground state is non-degenerate, Eqs.\ (\ref{wck})-(\ref{Cex}) are also
exactly valid for {\it finite} $n$ in the $T\rightarrow 0$ limit, using the
exact contractions
\begin{equation} g_L\equiv \langle c^\dagger_ic_j\rangle_0=
\frac{1}{n}\!\sum_{k\,{\rm occ.}}
 \cos(L\omega_k)\,, \label{gLn}\end{equation}
where $\langle O\rangle_0$ denotes ground state average and the sum runs over
the occupied levels (see next section).

\section{Results}
\subsection{Ground state transitions and concurrence}
Let us first describe the behavior in the $T\rightarrow 0$ limit. As $[H,N]=0$,
the ground state of $H$ can be characterized by the fermion number $N$, i.e.,
the total spin component $M=N-n/2$ in the spin representation. Since
$\lambda_k$ in (\ref{lk}) becomes {\it negative} for $b<v\cos\omega_k$, the
ground state {\it will exhibit $n$ transitions $N\rightarrow N+1$ as $b$
decreases from $|v|$ to $-|v|$}, starting from $N=0$ (the aligned state
$M=-n/2$) for $b>|v|$ ($\lambda_k>0$ $\forall$ $k$) and ending with  $N=n$
($M=n/2$) for $b<-|v|$ ($\lambda_k<0$ $\forall$ $k$).

For $v>0$, the first transition $0\rightarrow 1$ occurs at
\[b_1=v\,,\]
i.e., when the lowest negative parity level $\lambda_0=b-v$  becomes negative.
It represents for $b>0$ the {\it entangled-separable} border at $T=0$. For $b$
just below $b_1$ the ground state is the one-fermion state
${c'_0}^\dagger|0\rangle=\frac{1}{\sqrt{n}}\sum_jc^\dagger_j|0\rangle$, i.e.,
the $W$-state $(|\!\!\uparrow\downarrow\downarrow\ldots \downarrow\rangle+
|\!\!\downarrow\uparrow\downarrow\ldots\downarrow\rangle+\ldots)/\sqrt{n}$\,,
which exhibits {\it a constant concurrence}
 \begin{equation}C_{L}=2/n\,,\;\;\;\;(N=1)\label{CLs}\end{equation}
{\it for any separation $L$} (Fig.\ 1). Hence, the transition at $b=b_1$ is
from a fully separable state for $b>v$ (aligned state) to a state where {\it
any} pair is {\it equally entangled}.

\begin{figure}[t]
\vspace*{-1.8cm}

\centerline{\hspace*{1.cm}\scalebox{.7}{\includegraphics{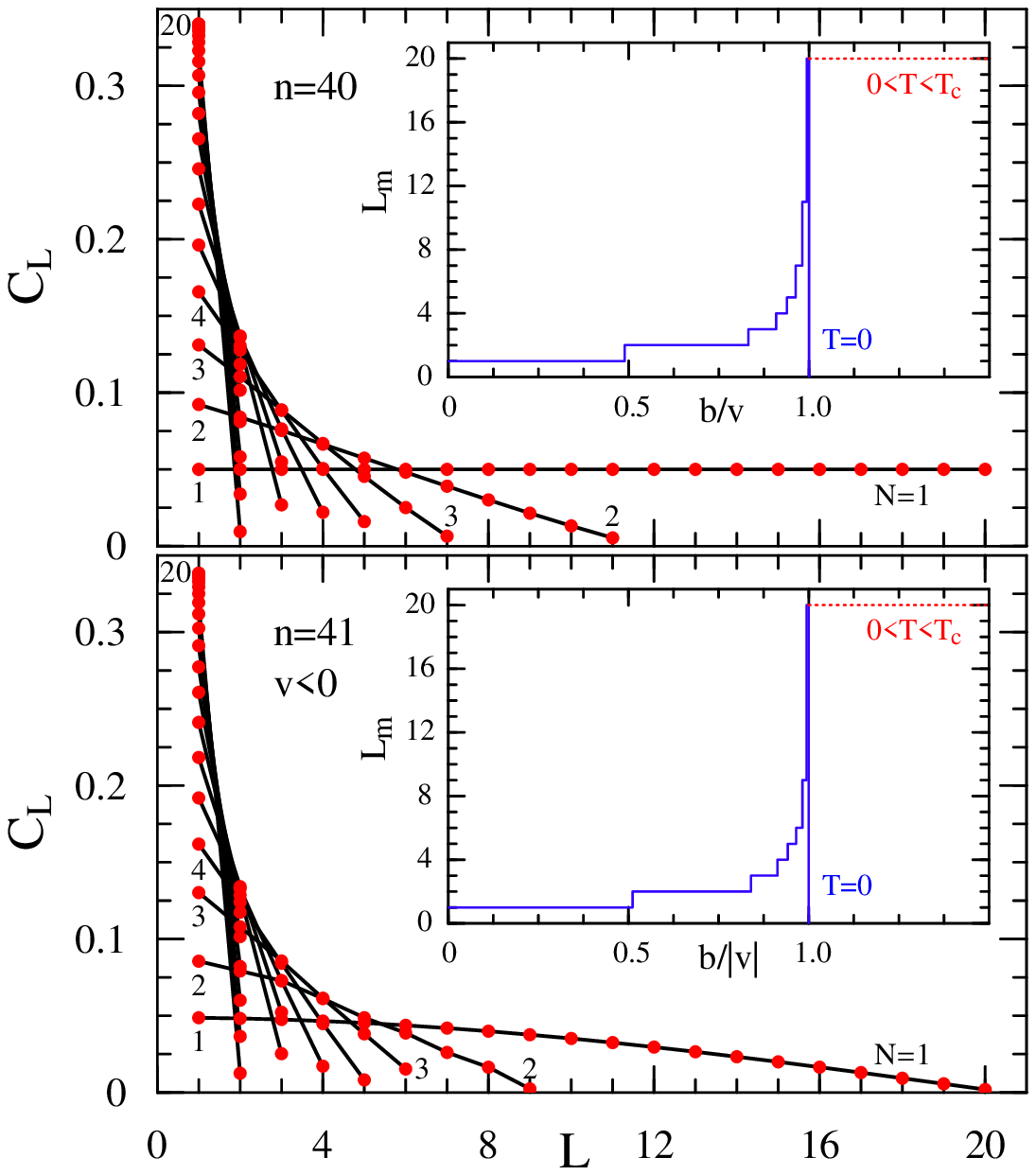}}}
 \vspace*{-10.2cm}

\caption{(Color online) Top panel: The concurrence $C_L$ as a function of site
separation $L$ in the 20 different entangled ground states of Hamiltonian
(\ref{H}) existing for $n=40$ qubits and $b>0$. Ground states are labelled by
the effective fermion number $N=M+n/2=1,\ldots,n/2$, which can be selected by
adjusting the magnetic field $b$ ($N$ fermions for $b_{N+1}<b<b_N$). For $N=1$,
$C_L$ has the same value for all separations (Eq.\ \ref{CLs}). The inset
depicts the entanglement range $L_m$ vs.\ the magnetic field in the same
system, which for $b>v$ vanishes at $T=0$ but remains maximum if $0<T<T_c\equiv
T_{n/2}$ (see next section). Bottom: Same details for an odd antiferromagnetic
chain with $n=41$ qubits (results for the $T\rightarrow 0$ limit of $\rho(T)$).
For $N=1$ all pairs are still entangled but $C_L$ decays for increasing $L$
(Eq.\ \ref{Cij2}).} \label{f1}\vspace*{-.3cm}
\end{figure}

Due to the parity dependence of the energy levels, the next transition
$1\rightarrow 2$ (existing for $n\geq 4$) does not take place when the next
$\lambda_k$ becomes negative ($b=v\cos\omega_k$) but rather when the lowest
$\sigma=1$ level crosses the previous $\sigma=-1$ level, i.e., when
$2\lambda_{\pm 1/2}=\lambda_0$, which leads to $b_2=v(2\cos(\pi/n)-1)$. In
general, for $v>0$ the transitions $N-1\rightarrow N$ occur at
$b_{N}=2v\cos\omega_k-b_{N-1}$, with $k=(N-1)/2$, which leads to the critical
fields
\begin{equation}
b_{N}=v\frac{\cos[(N-\half)\pi/n]}{\cos[\pi/(2n)]}\,, \;\;\;\;1\leq N\leq n\,,
\label{bcn}
\end{equation}
i.e., $b_N=v(\cos\omega_{k}-\sin\omega_k\tan(\pi/2n)$. Thus, $b_N<b_{N-1}$,
with $b_{n-N+1}=-b_{N}$ and $b_N\approx v\cos\omega_k$ for large $n$.

Eq.\ (\ref{CLs}) is valid for $b_2<b<b_1$. The exact expression for $C_L$ at
the other $N$-fermion ground states is given by Eq.\ (\ref{Cex}) with the
elements (\ref{gLn}), which become
\begin{equation}
 g_L=\frac{\sin (NL\pi/n)}{n\sin(L\pi/n)}\,, \label{gLn2}\end{equation}
with $g_0=N/n=\lim_{L\rightarrow 0} g_L$. For $N=1$, $g_L=1/n$ $\forall$ $L$
and Eq.\ (\ref{Cex}) leads to Eq.\ (\ref{CLs}).

For $N\geq 2$, $C_L$ will depend on the separation $L$, decreasing almost {\it
linearly} with $L$ for not too small $n$, as seen in Fig.\ \ref{f1}. A series
expansion of (\ref{Cex}) yields the initial trend $C_L\approx (2N/n)[1-\pi
L\sqrt{(N^2-1)/3}/n]$) for $NL\ll n$. The extent of pairwise entanglement
decreases then rapidly as $N$ increases (inset in Fig.\ \ref{f1}), the
separation between most distant entangled qubits being $L_m\approx
[(n+1.79)/3.57]$ for $N=2$ and, roughly, $L_m\approx[(n+4)/(2N)]$ for $2<N\ll
n/2$.

Just first and second neighbors ($L=1,2$) remain entangled for $|b|<0.65 v$
$\forall$ $n$ (and $|b|<0.82 v$ ($N\agt n/5$) for $n\rightarrow\infty$) whereas
only adjacent pairs ($L=1$) remain entangled for $|b|<0.26 v$ $\forall$ $n\neq
5$ (and $|b|<0.5 v$ ($N\agt n/3$) for $n\rightarrow \infty$) (for $n=5$ second
neighbors are entangled $\forall$ $b>0$). The concurrence of adjacent pairs
increases first linearly with $N$ ($C_1\approx 2N/n$ for $N\ll n$) and becomes
maximum for $N=n/2$ ($n>4$), where $g_1=1/(n\sin(\pi/n))\approx 1/\pi$ for
large $n$ and $C_1=2g_1(1+g_1)-1/2\approx 0.339$.

For odd $n$, results for $v<0$ must be separately examined. The lowest negative
parity level is now $\lambda_{\pm [n/2]}=b-|v|\cos(\pi/n)$, so that the first
transition occurs at
\[b_{1}^{-}=|v|\cos(\pi/n)\,,\;\;(v<0,n\;{\rm odd})\,,\]
with the ground state {\it two-fold degenerate} after the transition ($k=\pm
[n/2]$). The concurrence of the mixture $\half\sum_{k=\pm
[n/2]}|k\rangle\langle k|$ of the two degenerate ground states
$|k\rangle=\frac{1}{\sqrt{n}}\sum_{j}e^{-i\omega_k j}c^\dagger_j|0\rangle$ (the
$T\rightarrow 0$ limit of $\rho(T)$) is
\begin{eqnarray}
C_{L}^-&=& 2\cos(L\pi/n)/n\,,\;\;\;(N=1)
 \label{Cij2}\end{eqnarray}
{\it which is again non-zero $\forall$ $L$} ($n$ is odd) although it now {\it
decays} as $L$ increases (bottom panel in Fig.\ \ref{f1}).  For $L\ll n$,
$C^-_{L}\approx 2/n$, in agreement with (\ref{CLs}), whereas for most distant
qubits ($L=[n/2]$), $C^-_{L}=2\sin[\pi/(2n)]/n\approx \pi/n^2$.  Hence, for
large $L$  a significant odd-even difference in $C_{L}$ arises if $v<0$, {\it
even for large qubit number $n$}, due to the ground state degeneracy of the odd
system.

The next transition for $v<0$ and $n$ odd occurs when
$\lambda_{n/2}+\lambda_{n/2-1}=\lambda_{[n/2]}$, i.e., at
${b}_{2}^{-}=|v|[1+\cos(2\pi/n)]-{b}_{1}^{-}$,  and in general at
$b_N^-=|v|\sum_{k=N/2-1}^{N/2}\cos\omega_k-b_{N-1}^-$, which leads to the
smaller critical fields
\begin{equation}
b_N^{-}= b_N\cos(\pi/n),\;\;1\leq N\leq n\,, \label{bcn2}
\end{equation}
where $b_N$ are the fields (\ref{bcn}). Ground states remain two-fold
degenerate $\forall$ $N\neq 0,n$, since there is just one fermion in the
highest occupied level ($k=\pm (n-N)/2$).

Eq.\ (\ref{Cij2}) holds for $b_2^-<b<b_1^-$. The expression of $C_L^-$ for
general $N$ in the $T\rightarrow 0$ limit can be similarly obtained by using
Eqs.\ (\ref{Cex})-(\ref{gLn}) for each of the degenerate ground states and
taking then the average. The final result is
\begin{equation}C_L^-=\left[|{\rm Re}[{\rm Det}(A^-_L)]|
-2\sqrt{(g_0^2-g_L^2)((1-g_0)^2-g_L^2)}\right]_+
 \label{Cexi}\end{equation}
where $A_L^-$ is constructed with the elements
 \begin{equation}g_L^-=g_L e^{iL\pi/n}\,,\label{glnm}\end{equation}
with $g_L$ given again by (\ref{gLn2}). For $N=1$ (\ref{Cexi}) leads to Eq.\
(\ref{Cij2}). The behavior of $C_L^-$ for $N\geq 2$ is similar to that of $C_L$
(Eq.\ \ref{Cex}), although it is smaller than $C_L$ (due to the ground state
degeneracy) and its decay with $L$ is less linear (see bottom panel). For
instance, for $L=1$ ${\rm Re}({\rm Det} A_1^-)={\rm Det}(A_1)\cos(\pi/n)$,
whence $C_1^-<C_1$, with $C_1^-\rightarrow C_1$ for large $n$.

Let us finally mention that for $n\rightarrow \infty$ and  $\pi N/n\rightarrow
\omega$, with  $L$ finite, Eqs.\ (\ref{gLn2})-(\ref{glnm}) coincide both
exactly with the limit of (\ref{cijinf}) for $T\rightarrow 0$, where
$b_N\rightarrow v\cos\omega$.

\subsection{Results for finite temperatures}
\begin{figure}[t]
\vspace*{-0.2cm}

\centerline{\hspace*{-0.25cm}\scalebox{.5}{\includegraphics{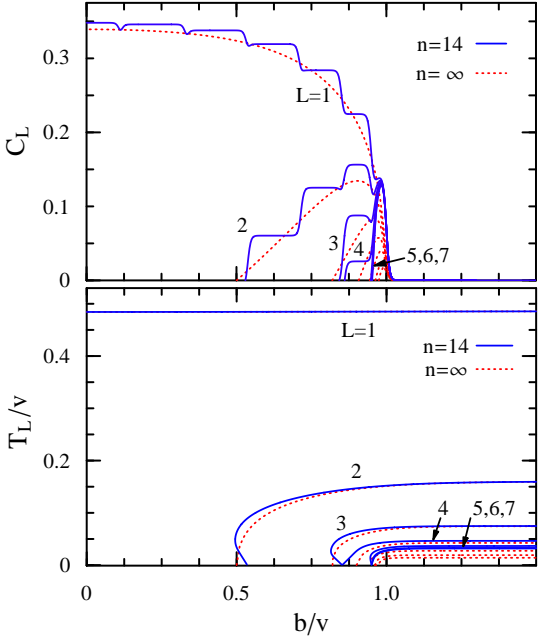}}}
 \vspace*{-0.5cm}

\caption{(Color online) Concurrence (top) and limit temperatures for
entanglement $T_L(b)$ (bottom) for pairs $i,i+L$ as a function of the magnetic
field, for $n=14$ qubits and in the thermodynamic limit $n\rightarrow\infty$.
The concurrence is plotted close to the $T\rightarrow 0$ limit ($T=0.005 v$).
All limit temperatures remain constant for $b/v\rightarrow\infty$ (see text).
Results for even $n$ lie mostly above those for $n\rightarrow\infty$,
particularly for large $L$ (where they saturate) and are independent of the
sign of $v$.} \label{f2}\vspace*{-0.1cm}
\end{figure}

\begin{figure}[t]
\vspace*{0.cm}

\centerline{\hspace*{-0.25cm}\scalebox{.5}{\includegraphics{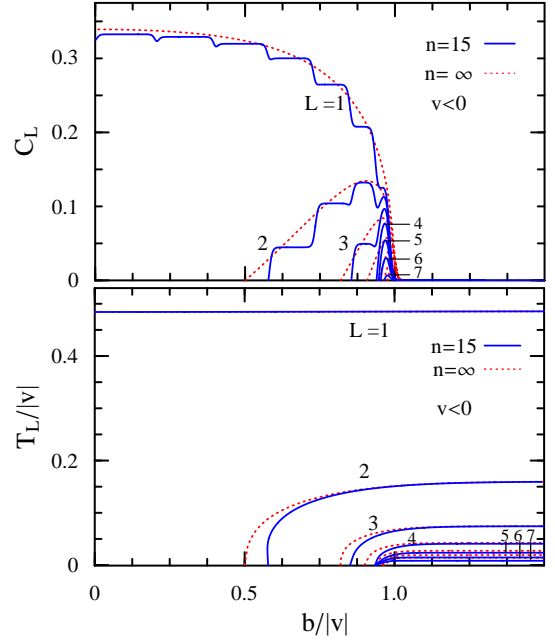}}}
 \vspace*{-0.5cm}

\caption{(Color online) Same details as Fig.\ \ref{f2} for $n=15$ qubits and
$v<0$. Results now lie mostly below those for $n\rightarrow\infty$ (which are
the same as in Fig.\ \ref{f2}) and do not saturate for large $L$.  Results for
$n=15$ and $v>0$ are similar to those of fig.\ \ref{f2}.}
 \label{f3}\vspace*{-0.4cm}
\end{figure}

Illustrative results for $n=14-15$ and the thermodynamic limit $n
\rightarrow\infty$ are depicted in Figs.\ \ref{f2}-\ref{f3}. For $T$ close to
$0$, the concurrence exhibits a stepwise behavior in finite chains, in
agreement with the $T=0$ transitions previously described, presenting dips at
the critical fields (\ref{bcn})-(\ref{bcn2}) due to the ground state degeneracy
at these points (level crossing). It is also verified that $C_L$ is smaller in
odd antiferromagnetic chains, particularly for large $L$ close to $n/2$, in
agreement with Eqs.\ (\ref{Cij2})-(\ref{Cexi}).

While at $T=0$ there is no entanglement in the ground state for $b>b_1$, a
fundamental result for $T>0$ is that $\rho(T)$ {\it remains entangled for all
fields $b>b_1$ if $T$ is sufficiently low}, leading to a small but {\it
non-zero} concurrence $C_{L}$ for {\it any} separation $L$ if $0<T<T_{L}(b)$.
Moreover, the limit temperature $T_{L}(b)$ approaches a {\it non-zero limit}
$T_L$ for $b\rightarrow\infty$ $\forall$ $L$, being practically constant for
$b\agt |v|$ (and $\forall b$ if $L=1$). This behavior applies for any $n$,
including $n\rightarrow\infty$ as well as the special case $v<0$ and $n$ odd,
as seen in the lower panels of Figs.\ \ref{f2}-\ref{f3}.

In order to rigorously prove the previous behavior, we note that for $b-|v|\gg
kT$ ($e^{-\beta(b-|v|)}\ll 1$), we may keep just zero, one and two fermion
states in $\exp[-\beta H]$, i.e.,
\[Z\approx e^{\beta bn/2}[1+\sum_{k\in K_{-}}\!\!
e^{-\beta \lambda_k}+\!\!\!\sum_{k<k'\in K_+}\!\!\!
 e^{-\beta(\lambda_k+\lambda_{k'})}]\]
and similarly for $\rho(T)$. This leads to $\alpha_L\approx e^{-\beta
b}I^-_L(\beta v)$ and $p^+_L\approx e^{-2\beta b} [{I_0^+}^2(\beta
v)-{I_L^+}^2(\beta v)]$ up to lowest order in $e^{-\beta b}$, where
\begin{eqnarray}
I_L^{\pm}(\beta v)&=&\frac{1}{n}\sum_{k\in K_\pm} e^{\beta
 v\cos\omega_k}\cos(L\omega_k)\,. \label{eqg}\end{eqnarray}
Hence, up to first order in $e^{-\beta b}$ we obtain
\begin{equation}
C_L\approx 2e^{-\beta b}\left[I_L^-(\beta v)-\sqrt{{I_0^+}^2(\beta v)-
{I_L^+}^2(\beta v)}\right]_+\,.
 \label{clt}\end{equation}
Thus, as $b$ increases the concurrence {\it decreases exponentially} with the
field when it is positive, but the limit temperature $T_L(b)$ becomes {\it
constant}, as the entanglement condition $C_L>0$ becomes $b$-independent, i.e.,
\begin{equation}{I^-_L}(\beta v)>\sqrt{{I^+_0}^2(\beta v)-{I^+_L}^2(\beta v)}
 \,. \label{inq}\end{equation}
Eq.\ (\ref{inq}) is {\it always satisfied} for sufficiently small but positive
$T$, for {\it any} distance $L$, ensuring  a non-zero concurrence and limit
temperature $T_L(b)$ for any $b>|v|$. This is easy to prove for $v>0$, where
for $T\rightarrow 0^+$, $I^-_L(\beta v)\approx e^{\beta v}/n>I^+_0(\beta
v)\approx 2e^{\beta v\cos(\pi/n)}/n$. It also holds for $v<0$ ($n$ odd), since
in this case,  for $T\rightarrow 0^+$, $I^-_L(\beta v)\approx e^{\beta
|v|\cos(\pi/n)}\cos(L\pi/n)/n$, whereas the r.h.s of (\ref{inq}) becomes
$\approx \sqrt{2}e^{\beta|v|\cos^2(\pi/n)}\sin(L\pi/n)/n<I_L^-(\beta v)$.

In the thermodynamic limit $n\rightarrow\infty$, and for {\it finite} $L\ll n$,
we may neglect parity effects and just replace
\begin{equation} I^{\sigma}_L(\beta v)\rightarrow \frac{1}{\pi}\int_{0}^\pi
 e^{\beta v\cos\omega}\cos(L\omega) d\omega= I_{L}(\beta v)\,,
 \end{equation}
where $I_L(x)$ is the modified Bessel function of the first kind
($I_L(x)\approx e^{x}[1+(1-4L^2)/8x]/\sqrt{2\pi x}$ for $x\rightarrow\infty$,
with $I_{L}(-x)=(-)^LI_L(x)$). Eq.\ (\ref{clt}) becomes then identical with the
result obtained from Eqs.\ (\ref{cijinf})-(\ref{Cex}) ($g_L\rightarrow
e^{-\beta b}I_L(\beta v)$ for $b-|v|\gg T$, with $ {\rm Det}(A_L)\rightarrow
2g_L$). Eq.\ (\ref{inq}) becomes then
 \begin{equation}\sqrt{2}I_L(\beta |v|)>I_0(\beta v)\,,\label{inqlt}
\end{equation}
which is again {\it always} satisfied for sufficiently low $T$ $\forall$ $L$.
The limit temperatures $T_L\equiv T_L(\infty)$ are then determined for
$n\rightarrow\infty$ by the equation $\sqrt{2}I_L(\beta|v|)=I_0(\beta|v|)$,
which  leads to $T_1\approx 0.486|v|$, $T_2\approx 0.16|v|$ and
 \begin{equation} T_L\approx |v|\ln 2/L^2\,,\label{ta1}\end{equation}
for large $L$ (as $I_L(x)/I_0(x)\approx e^{-L^2/(2x)}$ for $x\agt L^2$). Thus,
$T_L(b)$ {\it decreases as the inverse square of the pair distance $L$} for
large $b$. The maximum value attained by $C_L$ for $b>|v|$ becomes nevertheless
small and decays exponentially with both $b$ and $L^2$ ($C_L\approx
e^{-(b/|v|-1)L^2/t}f(t)/L$ for $T=|v|t/L^2<T_L$, with $f(t)=\sqrt{2t/\pi}
[e^{-t/2}-\sqrt{1-e^{-t}}]$). Eq.\ (\ref{ta1}) also indicates roughly the value
of $T_L(b)$ at the critical region $b\approx |v|$, since it is almost constant
for $b\agt |v|$ (Figs.\ \ref{f2}-\ref{f3}).

\begin{figure}[t]
\vspace*{-1.7cm}

\centerline{\hspace*{3.cm}\scalebox{.8}{\includegraphics{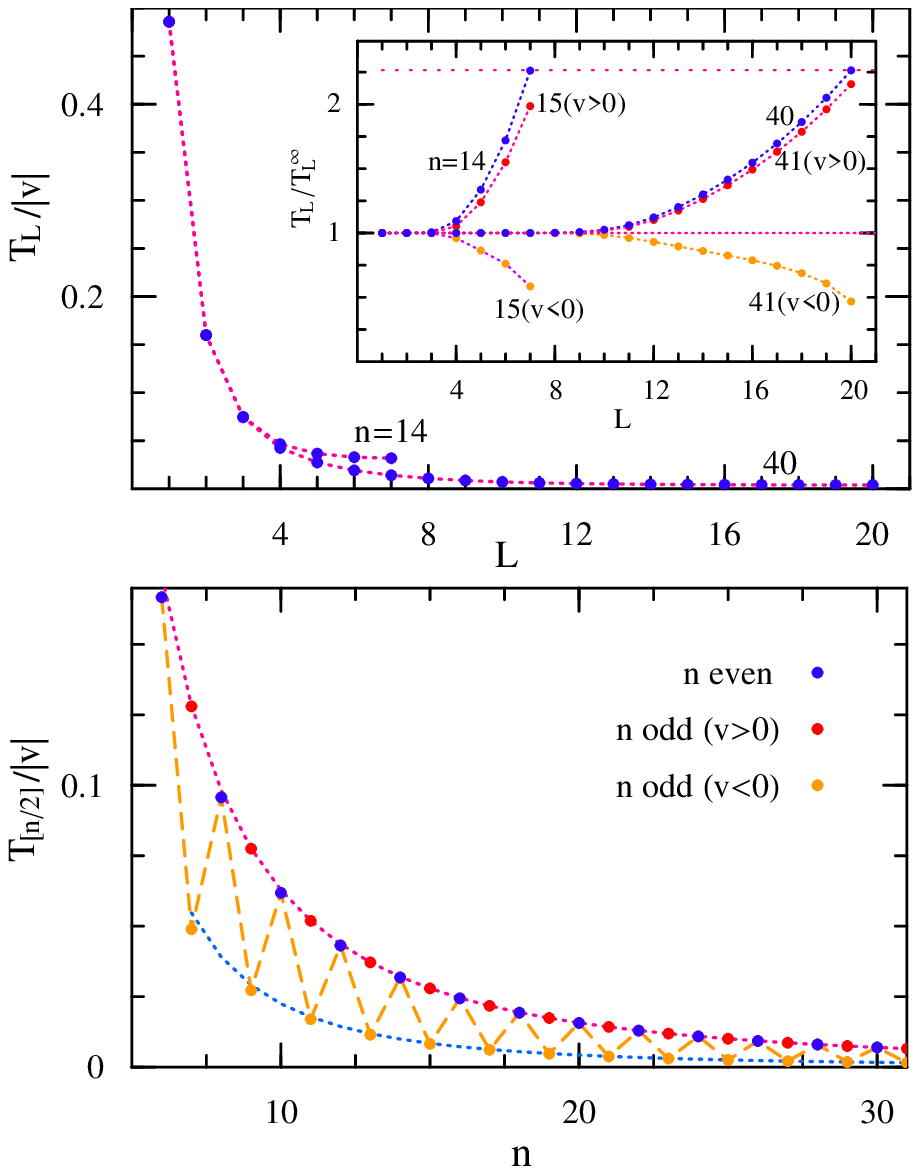}}}
 \vspace*{-12.8cm}

\caption{(Color online) Top: The limit temperatures $T_L$ for entanglement
between pairs for large magnetic fields $b\gg v$ as a function of separation
$L$, for $n=14$ and $n=40$ qubits, determined by Eq.\ (\ref{inq}). Inset: The
ratio between $T_L$ and the value in the thermodynamic limit $T_L^\infty$,
determined by Eqs.\ (\ref{inqlt})-(\ref{ta1}). $T_L$ deviates from $T_L^\infty$
for $L\agt n/4$, approaching, for $L\rightarrow n/2$, Eq.\ (\ref{ta2}) for $n$
even or $n$ odd and $v>0$, and Eq.\ (\ref{ta3}) for $n$ odd and $v<0$. Bottom:
The limit temperature for the most distant qubits ($L=[n/2]$), showing the
odd-even staggering arising for $v<0$ (dashed line). The upper (lower) dotted
line depicts the result of Eq.\ (\ref{ta2}) ((\ref{ta3})).}
 \label{f4}\vspace*{-0.4cm}
\end{figure}

On the other hand, for large $L\approx n/2$ the  projected  expression
(\ref{clt}) is required {\it even for large $n$}. For instance, for even $n$
and $L=n/2$, $\cos(L\omega_k)=0$ ($(-1)^k$) for $k$ half-integer (integer).
Hence, in this case $I^{+}_{n/2}(\beta v)=0$, while for $v>0$ and large $n$,
 \begin{equation}
I_{0(n/2)}^{+(-)}(\beta v)\approx e^{\beta v}\theta_{2(4)}
(e^{-2\beta v\pi^2/n^2})/n\,,
\end{equation}
after replacing $\cos\omega_k\approx 1-\omega_k^2/2$ ($\theta_2(u)\equiv
2\sum_{k=1/2}^\infty u^{k^2}$, $\theta_4(u)\equiv 1+2\sum_{k=1}^\infty
(-1)^ku^{k^2}$, denote the Elliptic Theta functions). These results also hold
approximately for large odd $n$ and $L=[n/2]$. Eq.\ (\ref{inq}) becomes then
$I^-_{n/2}(\beta v)>I^+_{0}(\beta v)$, and since $\theta_2(u)=\theta_4(u)$ for
$u=e^{-\pi}$, it leads to the limit temperature
\begin{equation} T_{[n/2]}\approx 2\pi v/\,n^2\,,\label{ta2}\end{equation}
for the most distant pairs and $v>0$. It is greater than Eq.\
(\ref{ta1}) for $L=n/2$ by a factor $\pi/(2\ln 2)\approx 2.27$.

Eq.\ (\ref{ta2}) does not hold for $v<0$ if $n$ is odd. In this case we may
directly employ the asymptotic expression of Eq.\ (\ref{inq}) for
$T\rightarrow 0^+$, which for large $n$ yields
\begin{equation}
T_{[n/2]}\approx \frac{|v|\pi^2}{2n^2\ln[2\sqrt{2}n/\pi]}\,,\;\;(v<0,\;n\;
 {\rm odd})\,.\label{ta3}\end{equation}
Thus, in this case there is an additional logarithmic factor in the
denominator, which makes $T_{[n/2]}$ {\it lower} than Eq.\ (\ref{ta2}) {\it and
also Eq.\ (\ref{ta1}) for $L=n/2$}, originating an odd-even staggering of
$T_{n/2}$ if $v<0$.

\begin{figure}[t]
\vspace*{-1.5cm}

\centerline{\hspace*{1.5cm}\scalebox{.7}{\includegraphics{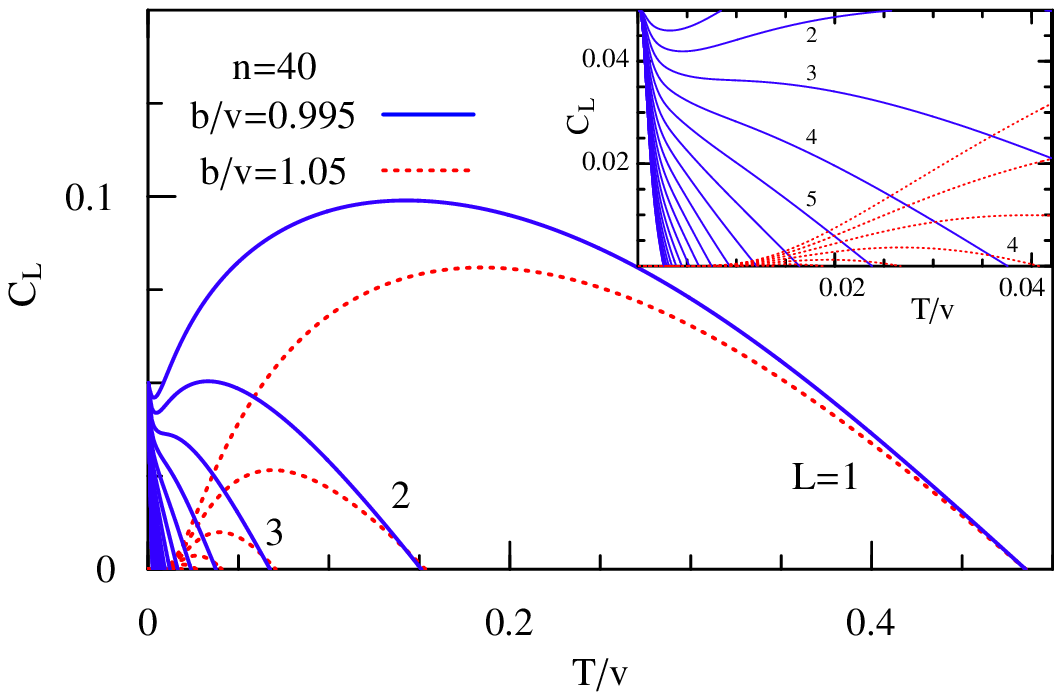}}}
 \vspace*{-15.cm}

\caption{(Color online) The thermal behavior of $C_L$ for $n=40$ qubits
near the transition at $b=v$. Limit temperatures  remain stable at the
transition, indicating the reentry of $C_L$ for $T>0$ $\forall$ $L$ for $b>v$.
The inset is an enlargement of the low $T$ region, showing the accumulation of
the limit temperatures for large $L$ at a value close to that given by Eq.\
(\ref{ta2}).}
 \label{f5}\vspace*{-0.2cm}
\end{figure}

The behavior of $T_L$ is depicted in Fig.\ \ref{f4}. It is seen that for $L\agt
n/4$, it deviates from the $1/L^2$ law given by Eq.\ (\ref{ta1}), approaching
the values given by Eqs.\ (\ref{ta2}) or (\ref{ta3}) for $L\approx n/2$. Fig.\
\ref{f5} depicts the typical thermal behavior of $C_L$ for $v>0$ near the
transition at $b=b_1$. For $b<b_1$ there is entanglement between all pairs if
$T$ is lower than a certain temperature, given approximately by Eq.\
(\ref{ta2}). It also shows the reentry of $C_L$ for $T>0$ for $b>v$, which is
quite prominent for low $L$.

\begin{figure}[t]
\vspace*{0cm}

\centerline{\hspace*{-0.25cm}\scalebox{.5}{\includegraphics{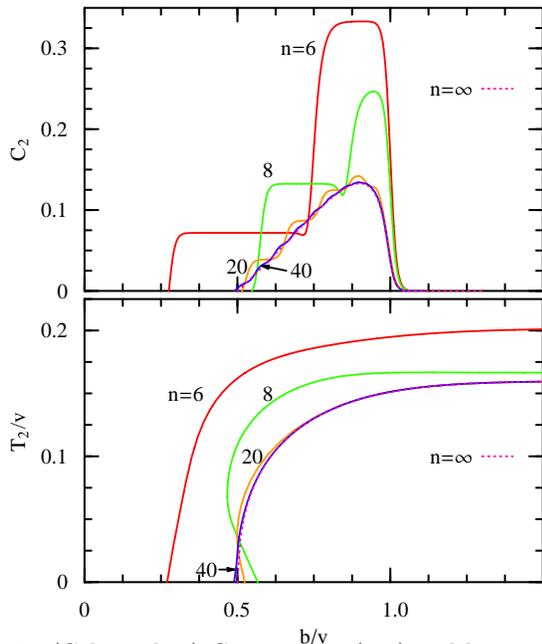}}}
 \vspace*{-0.8cm}

\caption{(Color online) Concurrence (top) and limit temperatures for
entanglement (bottom) for separation $L=2$ and different $n$, as a function of
the magnetic field. The concurrence is plotted at $T=0.01 v$. Dotted lines
depict the thermodynamic limit.}
\label{f6}\vspace*{-0.5cm}
\end{figure}

Finally Fig.\ \ref{f6} depicts the typical behavior with the qubit number $n$
of the concurrence and limit temperature. We have chosen a separation $L=2$.
Although the thermodynamic limit is on the average rapidly approached, the
stepwise behavior of $C_L$ at low $T\approx 0.01 v$ remains visible even for
$n=40$, and deviations in the limit temperature can be significant at the
onset. They are as well significant for small $n\alt 10$.

An interesting feature is that the slope of $T_L(b)$ can be {\it negative} in
this region, a fact already seen in Fig.\ \ref{f2} for $n=14$, and visible here
for $n=8$ and $n=20$. This occurs when the value of the onset field $b_c$ for
finite $n$ (which for $L=2$ corresponds to $b_2$, $b_3$, $b_7$ and $b_{14}$ for
$n=6,8,20$ and 40) lies {\it above} the value for $n\rightarrow\infty$, as
occurs for $n=8$, $20$. In these cases there is a small field interval below
$b_c$ where entanglement between second neighbors exists only above a {\it
threshold temperature} $T_L^i(b)>0$,  up to the higher limit temperature
$T_L(b)$.

A final comment is that we have checked all expressions by comparison with
calculations for low $n\alt 10$ based on the direct diagonalization of $H$. In
particular, for $n=2$, the entanglement condition (\ref{inq}) becomes exact
$\forall$ $b$, as in this case there are just one and two-fermion states,
reducing to $\sinh(\beta v)>1$ and leading to the known  limit temperature
$T_1=v/\ln(1+\sqrt{2})$ \cite{CR.04}.

\section{Conclusions}
We have provided an exact analytic treatment of the entanglement between
arbitrary pairs in cyclic $XX$ chains in the presence of a transverse magnetic
field, valid both at zero and finite temperatures and for {\it any} qubit
number $n$. We have shown that in spite of its simplicity, this system exhibits
very interesting features such as a discrete set of $[n/2]$ different entangled
ground states at $T=0$ (and $b>0$), which can be easily selected by adjusting
the magnetic field across the critical values (\ref{bcn}) or (\ref{bcn2}), and
which develop increasing entanglement ranges, reaching always full range (all
pairs entangled) in an interval $b_2<b<b_1$, even for odd antiferromagnetic
chains.

Moreover, while at $T=0$ the ground state is fully separable for $b>b_1$, we
have rigorously proved that for $T>0$ there is a small but non-zero
entanglement between {\it any} pair for all fields $b>b_1$ if $T$ is
sufficiently low, which decays exponentially with the field and with the square
of the separation $L$. Limit temperatures $T_L$ are roughly independent of $b$
for $b\agt v$ and decay as $L^{-2}$ for  $L\alt n/4$, but tend to saturate at
$T_{[n/2]}$ (Eq.\ \ref{ta2}) for  most distant pairs $(L\approx n/2$) if $n$ is
even or $v>0$. We have also shown that due to degeneracy of the ground state,
pairwise entanglement in odd antiferromagnetic chains is weaker, particularly
for distant pairs, where odd-even effects in the concurrence and $T_L$ subsist
for all $n$.

NC and RR acknowledge support of CONICET and CIC of Argentina.

\end{document}